\documentclass[12pt]{amsart}
\pdfoutput=1 

\usepackage{microtype}
\usepackage[utf8]{inputenc}
\usepackage[english]{babel}
\usepackage{fullpage}

\usepackage{csquotes}
\usepackage[sort,noadjust]{cite}

\usepackage[colorinlistoftodos,prependcaption,textsize=tiny]{todonotes}
\usepackage{siunitx}
\usepackage{soul}
\usepackage[svgnames]{xcolor}
\usepackage[draft=false]{minted}
\colorlet{CodeBackground}{black!5!white}
\setminted{encoding=utf8, mathescape, bgcolor=CodeBackground, 
  fontsize=\footnotesize} 

\usepackage{xspace}
\newcommand\Julia{\texttt{Julia}\xspace}
\newcommand\polymake{\texttt{polymake}\xspace}
\newcommand\OSCAR{\texttt{OSCAR}\xspace}
\newcommand\AbsGM{\texttt{GraphicalModel\{T,~L\}}\xspace}
\newcommand\GGM{\texttt{GaussianGraphicalModel\{T,~L\}}\xspace}
\newcommand\DGM{\texttt{DiscreteGraphicalModel\{T,~L\}}\xspace}
\newcommand\PM{\texttt{PhylogeneticModel\{T,~L\}}\xspace}
\newcommand\GPM{\texttt{GroupBasedPhylogeneticModel\{T,~L\}}\xspace}
\DeclareRobustCommand{\Macaulay}[1]{%
  \ifthenelse{\equal{\detokenize{#1}}{\detokenize{2}}}{}%
  {\PackageError{tboege-preprint}{Always use \protect\Macaulay2, never \protect\Macaulay\space alone}{}}%
  \texttt{Macaulay#1}\xspace}

\usepackage{pmboxdraw}
\DeclareUnicodeCharacter{2500}{\textSFx}

\usepackage[pdftex]{hyperref}
\hypersetup{
    colorlinks=true,
    linkcolor={magenta},
    citecolor={blue},
    urlcolor={blue}
}
\usepackage[nameinlink,capitalise]{cleveref}

\theoremstyle{plain}
\newtheorem{theorem}{Theorem}[section]

\theoremstyle{definition}
\newtheorem{remark}[theorem]{Remark}
\newtheorem{example}[theorem]{Example}

\newtheorem*{convention*}{Convention}

\newcommand{\cc}{\mathbb{C}}
\newcommand{\rr}{\mathbb{R}}

\newcommand{\cm}{\mathcal{M}}

\newcommand{\cg}{\mathcal{G}}
\newcommand{\ci}{\mathcal{I}}

\DeclareMathOperator{\rank}{rank}

\newcommand{\indep}{\mathrel{\text{$\perp\mkern-10mu\perp$}}}

\newcommand{\pd}{\mathrm{PD}}

\definecolor{benpurple}{RGB}{180, 0, 240}
\definecolor{antony}{RGB}{200, 20, 10}

\newcommand{\sym}{\mathrm{sym}}

\usepackage{graphicx} 

\title{\large Algebraic Statistics in {\Large\OSCAR}}

\author{Tobias Boege}
\address{
UiT -- The Arctic University of Norway}
\email{post@taboege.de}

\author{Antony Della Vecchia}
\address{Technical University of Berlin}
\email{vecchia@math.tu-berlin.de}

\author{Marina Garrote-López}
\address{Universitat Pompeu Fabra}
\email{marina.garrote@upf.edu}

\author{Benjamin Hollering}
\address{Max Planck Institute for Mathematics in the Sciences}
\email{benjamin.hollering@mis.mpg.de}

\date{\today}

\subjclass[2020]{%
  62R01, 
  62-04, 
  13P25  
  (primary)
  62H22, 
  92-08, 
  14-04  
  (secondary)%
}

\keywords{%
  algebraic statistics,
  algebraic phylogenetics,
  graphical model,
  computer algebra
}

\begin{document}

\begin{abstract}
We introduce the \texttt{AlgebraicStatistics} section of the \texttt{OSCAR} computer algebra system. We give an overview of its extensible design and highlight its features including serialization of data types for sharing results and creating databases, and state-of-the-art implicitization~algorithms.
\end{abstract}

\maketitle

\section{Introduction}
Algebraic statistics uses tools from algebraic geometry, commutative algebra, and combinatorics to address problems in statistics. This approach is based on the insight that, in many reasonable and practically useful instances, a parametric statistical model~$\cm$ is given as the image $\varphi(\Theta)$ of a ``simple'' semialgebraic set~$\Theta$ under a polynomial (or rational) map~$\varphi$. Many~interesting statistical properties of $\cm$ are captured by the Zariski closure $V = \overline{\cm} = \overline{\varphi(\Theta)}$ which can be studied using computer algebra methods like Gröbner~bases.
Since the seminal paper of Diaconis and Sturmfels~\cite{DiaconisSturmfels}, which forged a surprising connection between Fisher's exact test and Gr\"obner bases of toric ideals, algebraic methods have found numerous new applications in statistics. These include conditional independence implication \cite{GaussianBN,Fink,CIHidden,MatusRings,BoegeDiss,LyapCI}, structure and parameter identifiability of statistical models \cite{Allman2009Identif, Allman2024, ColoredDAG}, causal discovery \cite{Drton2025,wang2019highdimensional}, and maximum likelihood estimation \cite{MLdegree,MLGaussian,MLBrownian}.
For a detailed dive into algebraic statistics and the various ways in which algebraic tools have been employed to solve statistical problems we refer the reader to~\cite{Sullivant}. 

This paper presents the \texttt{AlgebraicStatistics} section of the computer algebra system \OSCAR \cite{OSCAR-book, OSCAR} which first appeared in version 1.2.0. It provides algebraic solutions to a range of statistical tasks. The current functionality is centered around \emph{graphical models} and \emph{phylogenetics}. 
Using the type system of the \Julia programming language, we created a framework which makes it easy for the user or future developers to build new families of models on top of our existing functionality.

The \Julia ecosystem is steadily growing and connects \OSCAR with sophisticated visualization, networking, database and numerical algebraic geometry toolkits --- and more.
Readers looking to run the examples discussed in this paper can find a Jupyter notebook at \url{https://github.com/dmg-lab/OscarAlgebraicStatistics}.

\section{The \texorpdfstring{\normalfont\mintinline[fontsize=]{julia}{typeof}}{'typeof'} Graphical Models}

Graphical models are statistical models derived from graphs. The vertices of the graph correspond to random variables and the edges limit the influence of some random variables on others, by prescribing sparsity patterns in the parametrization. In this way, the graph affects qualitative features of the distributions in the model, usually by imposing conditional independence constraints. Thanks to their interpretability, graphical models have seen widespread use in applications across the sciences: biologists seeking to organize the lineage of species into phylogenetic networks~\cite{Phylogenetics,MathBiology}, econometrists~\cite{EconCausal} and social scientists~\cite{SocialCausal} for causal modeling, engineers for computer vision~\cite{VisionGraphs}, or environmental scientists who produced a concrete graphical model of the California water reservoirs~\cite{Water}. This has resulted in broad interest in their theoretical foundations; see~\cite{GraphicalModels,Lauritzen,Pearl}.

In general, a graphical model is a parametric statistical model which is given by the image of a rational map $\varphi_\cg\colon N \to M$ whose details depend on a graph~$\cg$ and the choice of distribution type:
\begin{itemize}
\item For Gaussian random variables, $\Sigma = \varphi_\cg(\theta) \in \pd_n$ is the covariance matrix.
\item For $n$ discrete random variables with $d_1, \dots, d_n \ge 1$ states, $p = \varphi_\cg(\theta) \in \Delta(d_1, \dots, d_n)$ is the non-negative $d_1 \times \cdots \times d_n$ tensor of atomic probabilities.
\end{itemize}
The graph~$\cg$ can be of many different types, such as undirected or directed, each of which yields a different parametrization $\varphi_\cg$ and a different model.

We implement several commonly studied families of graphical models as instances of the abstract parametric type \AbsGM. The main idea of this framework is to allow users to implement their own graphical model types while reusing as much of the available code as possible. 
Users can create their own graphical model $\cm = \varphi_\cg(N) \subseteq M$,  given by the image of a rational map $\varphi_\cg\colon N \to M$, by overloading the appropriate attribute functions
\begin{itemize} 
\item \texttt{parameter\_ring} returns the coordinate ring of the parameter space $N$,
\item \texttt{model\_ring} is the coordinate ring of the ambient space $M$,
\item \texttt{parametrization} is the ring homomorphism from \texttt{model\_ring} to \texttt{parameter\_ring} which is the pullback of the parametrization $\varphi_\cg$.
\end{itemize}

We will show an example of this in \Cref{example:cgm}.

\Julia's type system allows for sophisticated method dispatch functionality; we refer the interested reader to \cite[\S3, \S4]{Julia-2017} and note here regarding terminology that a \emph{function} in \Julia is a name referring to a collection of type-specific \emph{methods}. Calling \texttt{parametrization(M)} invokes the most appropriate method of the function \texttt{parametrization} for handling the argument type of~\texttt{M}. Thus, we can provide a unified interface which can accommodate the whole ``spectrum of graphical models''. 
In \Cref{sec:vanishing ideal} we show where \Julia's multiple dispatch philosophy really shines: we can transparently provide faster algorithms for certain types of graphs for which there is a theorem enabling computational shortcuts.

Following the Gaussian vs.\ discrete dichotomy, we provide the following concrete~types
\begin{itemize}
\item \GGM and \DGM;
\item \PM and \GPM.
\end{itemize}
An example Gaussian graphical model can be created like this:
\begin{minted}{julia-repl}
julia> G = graph_from_edges(Undirected, [[1,2], [1,4], [2,3], [3,4]])
Undirected graph with 4 nodes and the following edges:
(2, 1)(3, 2)(4, 1)(4, 3)
\end{minted}
\clearpage
\begin{minted}{julia-repl}
julia> M = gaussian_graphical_model(G)
Gaussian Graphical Model on a Undirected graph with 4 nodes and 4 edges

julia> typeof(M)
GaussianGraphicalModel{Graph{Undirected}, Nothing}
\end{minted}
Creating \texttt{M} from the undirected graph \texttt{G} infers the type parameter \mintinline{julia}{T = Graph{Undirected}}. Possible values for \texttt{T} include but are not limited to:
\begin{itemize}
\item \texttt{Graph\{Directed\}}, \texttt{Graph\{Undirected\}}, or \texttt{MixedGraph}; 
\item \texttt{PhylogeneticTree} or \texttt{PhylogeneticNetwork}. 
\end{itemize}
The labeling type \texttt{L} is a \texttt{NamedTuple} type from \Julia's standard library. It can be used to attach additional data to the graph via a \texttt{GraphMap} from \polymake. \Cref{example:cgm} shows how to use this to implement \emph{colored} Gaussian graphical models as introduced in~\cite{ColoredGraphs}.

\subsection{Gaussian Graphical Models on Undirected Graphs}

The object \texttt{M} from the above \Julia session represents an undirected Gaussian graphical model on four vertices. This model is a semialgebraic set constructed as follows. Let $\cg = ([n], E)$ be an undirected graph and let $\mathcal L_G = \{K \in \cc^{n \times n}_\sym ~:~ k_{ij} = 0 \text{ for } (i, j) \in E(G)\}$ \} be the linear space of symmetric matrices supported on the edges of $\cg$. Then the Gaussian graphical model corresponding to $\cg$ is $\cm_\cg = \{\Sigma \in \rr^{n \times n} ~:~ \Sigma^{-1} \in \mathcal L_G \} \cap \pd_n$ which is the intersection of an inverse linear space with $\pd_n$. The model $\cm_\cg$ can also be seen as the image of the rational map
\begin{equation}
\label{eq:UndirectedParam}
\begin{aligned}
\varphi_\cg: \mathcal L_G \cap \pd_n \to \pd_n \\
                 K \mapsto K^{-1}.
\end{aligned}
\end{equation}
In algebraic statistics, one often ignores the constraint that $\Sigma \in \cm_\cg$ be positive definite and instead studies the algebraic closure of the model, $V_\cg = \overline{\cm_\cg}$, which is parametrized by extending the rational map $\varphi_\cg$ to all of $\mathcal L_G$. This is justified on a case-by-case basis by observing that statistical properties of interest only depend on~$V_\cg$. From an algorithmic point of view, it is much easier to work with the variety $V_\cg$ than with the semialgebraic set $\cm_\cg$.

This graphical model can also be specified as the set of positive definite matrices which satisfy a collection of conditional independence (CI) statements which is known as the \emph{global Markov property} of $\cg$. Since a Gaussian random vector satisfies the conditional independence $A \indep B \mid C$ if and only if $\rank(\Sigma_{A \cup C, B \cup C}) < \# C + 1$, the set of multivariate Gaussian distributions which satisfy a collection of CI statements $\mathcal{C}$ can also be seen as the intersection of a variety $V_\mathcal{C} = V(I_\mathcal{C}) = \{\Sigma \in \cc^{n \times n}_\sym ~:~ \rank(\Sigma_{A \cup C, B \cup C}) < \# C + 1 \text{ for all } A \indep B \mid C \in \mathcal{C} \}$ with $\pd_n$. The user can compute the \emph{conditional independence ideal} $I_{\mathcal C}$ as well as the vanishing ideal $\ci(V_\cg)$ of the model. For instance, in our running example:
\begin{minted}{julia-repl}
julia> C = global_markov(graph(M))
2-element Vector{CIStmt}:
 [1 _||_ 3 | {2, 4}]
 [2 _||_ 4 | {1, 3}]

julia> J = ci_ideal(gaussian_ring(M), C)
Ideal generated by
  -s[1, 2]*s[2, 3]*s[4, 4] + s[1, 2]*s[2, 4]*s[3, 4] + ...
  -s[1, 1]*s[2, 3]*s[3, 4] + s[1, 1]*s[2, 4]*s[3, 3] + ...

\end{minted}
\clearpage
\begin{minted}{julia-repl}
julia> I = vanishing_ideal(M)
Ideal generated by
  -s[1, 2]*s[2, 3]*s[4, 4] + s[1, 2]*s[2, 4]*s[3, 4] + ...
  -s[1, 1]*s[2, 3]*s[3, 4] + s[1, 1]*s[2, 4]*s[3, 3] + ...
  s[1, 1]*s[2, 3]*s[2, 4]*s[3, 4] - s[1, 1]*s[2, 4]^2*s[3, 3] ...
  -s[1, 1]*s[2, 2]*s[2, 3]*s[3, 4]^2 + s[1, 1]*s[2, 2]*s[2, 4]*s[3, 3]*s[3, 4] ...
\end{minted}
These two ideals are generated differently and are not in general the same but for the graph~\texttt{G} they happen to coincide as can be verified by running \mintinline{julia}{I == J}.

\subsubsection{Introducing Colors}
\label{example:cgm}

In this section we show by example how our framework supports the implementation of new types of graphical models.
A colored (undirected, Gaussian) graphical model according to \cite{ColoredGraphs} is obtained by restricting the linear space $\mathcal L_\cg$ further. The graph $\cg$ is augmented by a map $c$ which assigns a color to each vertex and edge. We define a linear space $\mathcal L_{\cg,c}$ of symmetric matrices $K$ via:
\begin{itemize}
\item $k_{ij} = 0$ if $ij \not\in E$,
\item $k_{ij} = k_{st}$ if $ij, st \in E$ and $c(ij) = c(st)$, and
\item $k_{ii} = k_{ss}$ if $i, s \in V$ and $c(i) = c(s)$.
\end{itemize}
The parametrization map from \eqref{eq:UndirectedParam} is restricted to $\mathcal L_{\cg,c}$ but stays otherwise the same.
In~\Julia we create the edge and vertex colors as hashmaps (\texttt{Dict} in \Julia) and pass them to the \texttt{graph\_from\_labeled\_edges} function along with the type of graph and name for the labeling.

\begin{minted}{julia-repl}
julia> edge_labels = Dict((1, 4) => "Green", (2, 3) => "Green",
                          (3, 4) => "Blue", (1, 2) => "Blue");

julia> vertex_labels = Dict(1 => "Red", 2 => "Red", 3 => "Yellow", 4 => "Yellow");

julia> G = graph_from_labeled_edges(Undirected, edge_labels, vertex_labels; name=:color);

julia> Mc = gaussian_graphical_model(G);

julia> typeof(Mc)
GaussianGraphicalModel{Graph{Undirected}, @NamedTuple{color::Oscar.GraphMap{Undirected}}}

julia> Ic = vanishing_ideal(Mc)
Ideal generated by
  -s[1, 2]*s[2, 3]*s[4, 4] + s[1, 2]*s[2, 4]*s[3, 4] + ...
  -s[1, 1]*s[2, 3]*s[3, 4] + s[1, 1]*s[2, 4]*s[3, 3] + ..
  s[1, 1]*s[2, 3]*s[2, 4]*s[3, 4] - s[1, 1]*s[2, 4]^2*s[3, 3] ...
  -s[1, 1]*s[2, 2]*s[2, 3]*s[3, 4]^2 + s[1, 1]*s[2, 2]*s[2, 4]*s[3, 3]*s[3, 4] ...
\end{minted}

The snippet above shows that passing an undirected graph labeled with \texttt{color} to the \texttt{gaussian\_graphical\_model} function yields a \GGM object with 
  \mintinline{julia}{L = @NamedTuple{color::Oscar.GraphMap{Undirected}}}.
However, the \texttt{vanishing\_ideal} is odd: it is the same ideal as for the uncolored model!

This is because \OSCAR 1.7.0 does not have methods built in to work with the \texttt{color} labeling. As a result, the general method for \texttt{vanishing\_ideal} will be invoked, which does not take the \texttt{color}s into account. To~teach \OSCAR about colored graphical models, we only need to overload \texttt{parameter\_ring}:
\clearpage
\begin{minted}[linenos]{julia}
import Oscar: parameter_ring, GraphDict, GraphMap

const ColoredGGM{Undirected} = GaussianGraphicalModel{
  Graph{Undirected}, @NamedTuple{color::GraphMap{Undirected}}
}

@attr Tuple{
  QQMPolyRing,
  GraphDict{QQMPolyRingElem}
} function parameter_ring(GM::ColoredGGM{Undirected})
  G = graph(GM)
  colors = unique([[G.color[e] for e in edges(G)];
                   [G.color[v] for v in vertices(G)]])
  R, x = polynomial_ring(QQ, varnames(GM)[:k] => colors)
  color_dict = Dict{String, MPolyRingElem}(
    color => x[i] for (i, color) in enumerate(colors))

  gens_dict = GraphDict{QQMPolyRingElem}(
    Dict{Union{Int, Edge}, QQMPolyRingElem}(merge(
      Dict(e => color_dict[G.color[e]] for e in edges(G)),
      Dict(v => color_dict[G.color[v]] for v in vertices(G))
      )))
  return R, gens_dict
end
\end{minted}

We break down this snippet line by line.
Line 1 imports the function \texttt{parameter\_ring} and some internal types from \OSCAR.
Explicitly importing the function is necessary to extend it with a new method (overloading), while importing the types makes the code a little more readable.
Lines 3--5 define an alias for our colored gaussian model type (\texttt{ColoredGGM\{Undirected\}}) to enhance readability.
The core method definition begins on line 7. The \texttt{@attr} macro is used so that the parameter ring is only computed once per model and then cached.
The~function signature, defined on lines 7--10, reads as follows: given a colored gaussian model on an undirected graph the \texttt{parameter\_ring} returns a tuple with first, a multivariate polynomial ring over $\mathbb{Q}$, and second, a mapping (hashmap) from edges and vertices to generators of the ring.
Lines 11--14 create the polynomial ring necessary to represent each unique color on the graph as a variable.
Note that if \texttt{color} is a labeling on the graph~\texttt{G} then it can be accessed as a property, i.e., \texttt{G.color[e]} will return the color of the edge~\texttt{e}.
The~remainder of the snippet creates a mapping from each color to the variable in the polynomial ring (lines 15--16) and then uses it to create a mapping from the edges and vertices of the graph to their respective variables in the polynomial ring, given their color (lines 18--23). 

Whenever a graphical model now involves a labeling with the name \texttt{color}, \Julia's multiple dispatch will prefer the above specialized version of \texttt{parameter\_ring}. The \texttt{parametrization} function $K \mapsto K^{-1}$ and the \texttt{model\_ring} remain the same in the colored model, so we do not have to overload them; just as in the mathematical definition, it suffices to restrict the domain of the parametrization map via \texttt{parameter\_ring}. The default \texttt{vanishing\_ideal} method merely computes the kernel of \texttt{parametrization} and thus automatically becomes~correct,~too:
\begin{minted}{julia-repl}
julia> include("/path/to/overloaded/function/file.jl");

julia> Mc = gaussian_graphical_model(G);

julia> Ic = vanishing_ideal(Mc)
Ideal generated by
  s[3, 3] - s[4, 4]
  s[1, 4] - s[2, 3]
  s[1, 3] - s[2, 4]
  s[1, 1] - s[2, 2]
  s[1, 2]*s[2, 3] - s[2, 2]*s[2, 4] ...
  s[1, 2]*s[2, 4]*s[3, 4] - ...
  s[1, 2]^2*s[3, 4] - s[1, 2]*s[2, 4]^2 - ...
  -s[1, 2]^3*s[3, 4] + s[1, 2]^2*s[2, 4]^2 + ...
\end{minted}
Compare the outputs of this example with \cite[Examples~2.2 and~2.3]{GoebelMisra25}.

\subsection{Phylogenetics}

Phylogenetics provides another canonical example of algebraic statistical models. Its goal is to reconstruct the evolutionary history of biological entities based on observed genetic data (e.g., DNA sequences). This history is represented by a \emph{phylogenetic tree} $\mathcal{T}=(V, E)$ (a directed acyclic graph) rooted at an interior node $r$ (or, more generally, a \emph{phylogenetic network} which we defer to \Cref{sec:phylo network}). We model the evolution of genetic states as a Markov process on $\mathcal{T}$. To each node $v \in V$, we associate a random variable $X_v$ taking values in a state space of nucleotides $ \{\texttt A, \texttt C, \texttt G, \texttt T\}$.

In this framework, the {parametrization} of this model is defined by the stochastic parameters of the Markov process. The parameter space $\Theta$ consists of a probability distribution $\pi$ at the root and a set of $4 \times 4$ transition matrices $\mathcal{M} = \{M_e\}_{e \in E}$. For every directed edge $e = u \to v$, the entry $M_e(x, y) = P(X_v = y \mid X_u = x)$ represents the probability of a state $x$ at the parent node $X_u$ substituting to state $y$ at the child node $X_v$. When no constraints are placed on $\Theta$, this is known as the \emph{General Markov model}. Classical biological models (e.g., \emph{Jukes--Cantor}, \emph{Kimura}, \emph{GTR}) correspond to sub-models where $\Theta$ is restricted by linear constraints reflecting biochemical properties.

The \emph{phylogenetic model} $\mathcal{M}_\mathcal{T}$ corresponds to the set of probability distributions observable at the leaves of $\mathcal{T}$ (labeled by $[n]$). It is a discrete, directed graphical model with \emph{hidden variables}. Under the local Markov property, the joint distribution on the tree factors over the edges. However, because internal nodes represent latent (unobserved) ancestral species, the probability of observing a specific pattern $i_1 \dots i_n$ at the leaves is obtained by marginalizing over the internal states. This defines the polynomial map $\varphi_\mathcal{T}: \Theta \to \mathcal{M}_\mathcal{T}$, where the coordinate functions are given by
\begin{equation*}
    p_{i_1 \dots i_n} = \sum_{\substack{x_v \in  \{\texttt A, \texttt C, \texttt G, \texttt T\},\\ v \text{ internal}}} \left( \pi(x_r) \prod_{u \to v \in E} M_e(x_u, x_v) \right).
\end{equation*}
Thus, the phylogenetic model $\mathcal{M}_\mathcal{T}$ is the image of the parameter space $\Theta$ under the map~$\varphi_\mathcal{T}$. Algebraic statistics allows us to study $\mathcal{M}_\mathcal{T}$ by analyzing the \emph{phylogenetic invariants} --- polynomials in the leaf probabilities that vanish on $\mathcal{M}_\mathcal{T}$ --- which generate the ideal defining the Zariski closure of the model.
These invariants are crucial tools for distinguishing between different tree topologies without any prior knowledge.

These structures can be constructed and manipulated via the \texttt{PhylogeneticModel} type. This object encapsulates the tree topology, the model type, and the underlying polynomial~ring.

\begin{example}\label{ex:JCprob}
    In this example, we construct a simple tree with 3 leaves and assign a Jukes--Cantor structure to the transition matrices. Note that while we define a generic matrix structure with variables $a$ and $b$, the \texttt{phylogenetic\_model} function automatically assigns distinct parameters ($a_i, b_i$) to each edge $i$, allowing for varying branch lengths.

\begin{minted}{julia-repl}
julia> T = phylogenetic_tree(graph_from_edges(Directed, [[4,1], [4,2], [4,3]]))
Phylogenetic tree with QQFieldElem type coefficients

# Define the Jukes--Cantor substitution structure
# a: probability of remaining same, b: probability of mutation
julia> M_JC = [:a :b :b :b;
               :b :a :b :b;
               :b :b :a :b;
               :b :b :b :a];

julia> PM = phylogenetic_model(T, M_JC)
Phylogenetic model on a tree with 3 leaves and 3 edges 
  with root distribution [1//4, 1//4, 1//4, 1//4] and transition matrices of the form ...

julia> parameter_ring(PM)[1]
Multivariate polynomial ring in 6 variables a[1], a[2], a[3], b[1], ..., b[3]
  over rational field

julia> entry_transition_matrix(PM, 3, 3, Edge(4, 1))
a[1]

julia> entry_transition_matrix(PM, 1, 2, Edge(4, 2))
b[2]
\end{minted}

The function \texttt{model\_ring} generates the ambient ring for the leaf probabilities, while \texttt{parametrization} returns the polynomial map $\varphi_\mathcal{T}$. Finally, the phylogenetic invariants defining the model are computed using \texttt{vanishing\_ideal}: 

\begin{minted}{julia-repl}
julia> Rp, p = model_ring(PM); p
Dict{Tuple{Vararg{Int64}}, QQMPolyRingElem} with 5 entries:
  (1, 1, 1) => p[1,1,1]
  (1, 2, 1) => p[1,2,1]
  ...

julia> phi = parametrization(PM);

# Compute the probability of observing state A (index 1) at all leaves
julia> phi(p[(1, 1, 1)]) 
1//4*a[1]*a[2]*a[3] + 3//4*b[1]*b[2]*b[3]

julia> vanishing_ideal(PM)
Ideal generated by
  2p[1,2,3]^3 - p[1,2,3]^2p[1,2,2] - p[1,2,3]^2p[1,2,1] - p[1,2,3]^2p[1,1,2] + 
  p[1,2,3]^2p[1,1,1] - p[1,2,3]p[1,2,2]^2 - p[1,2,3]p[1,2,1]^2 - p[1,2,3]p[1,1,2]^2 + 
  p[1,2,3]p[1,1,1]^2 + p[1,2,2]^2p[1,2,1] + p[1,2,2]^2p[1,1,2] + p[1,2,2]p[1,2,1]^2 - 
  p[1,2,2]p[1,2,1]p[1,1,2] - p[1,2,2]p[1,2,1]p[1,1,1] + p[1,2,2]p[1,1,2]^2 - 
  p[1,2,2]p[1,1,2]p[1,1,1] + p[1,2,1]^2p[1,1,2] + p[1,2,1]p[1,1,2]^2 - 
  p[1,2,1]p[1,1,2]p[1,1,1]
\end{minted}
\end{example}
\OSCAR includes a number of common biological models, such as
\begin{itemize}
\item group-based models \texttt{jukes\_cantor\_model}, \texttt{kimura2\_model}, and \texttt{kimura3\_model},
\item \texttt{general\_markov\_model}, and \texttt{general\_time\_reversible\_model}.
\end{itemize}

\subsubsection{Group-based Models}
\label{sec: group-based}

The \emph{group-based models} (e.g., Jukes--Cantor, Kimura 2 or 3 parameters) admit a specific linear change of coordinates via the discrete Fourier transform. This transformation diagonalizes the transition matrices and converts the probability coordinates into \textit{Fourier coordinates} (denoted by $q_{i_1 \dots i_n}$). In this new coordinate system, the parametrization map $\psi_\mathcal{T}$ simplifies to a monomial map in terms of the \textit{Fourier parameters} (the eigenvalues of the transition matrices). Algebraically, this identifies the model as a \textit{toric variety}, a structure that significantly simplifies computational tasks, see~\cite{Sullivant} for more details.

To utilize this structure, we provide the \texttt{GroupBasedPhylogeneticModel} type. Instances are constructed using specific functions like \texttt{jukes\_cantor\_model(T)} and contains an underlying \texttt{PhylogeneticModel} object, which remains accessible via the \texttt{phylogenetic\_model} function.
Furthermore, the explicit linear transformation from Fourier to probability coordinates is obtained via \texttt{coordinate\_change}, with the reverse mapping provided by \texttt{inverse\_coordinate\_change}.

\begin{example}
    This example presents the model defined in \Cref{ex:JCprob} utilizing the \texttt{GroupBasedPhylogeneticModel} type, where computations are performed directly in terms of Fourier parameters and Fourier coordinates.
    \begin{minted}{julia-repl}
julia> PM = jukes_cantor_model(T) # Define a Jukes--Cantor model on the tree T
Group-based phylogenetic model on a tree with 3 leaves and 3 edges 
  with root distribution [1//4, 1//4, 1//4, 1//4], 
    transition matrices of the form ...
  and fourier parameters of the form [:x, :y, :y, :y].

julia> parameter_ring(PM)[1]
Multivariate polynomial ring in 6 variables x[1], x[2], x[3], y[1], ..., y[3]
  over rational field

julia> Rq, q = model_ring(PM); q
Dict{Tuple{Vararg{Int64}}, QQMPolyRingElem} with 5 entries:
  (1, 1, 1) => q[1,1,1]
  (2, 3, 4) => q[2,3,4]
  ···

julia> entry_fourier_parameter(PM, 3, Edge(4, 1))
y[1]

julia> entry_fourier_parameter(PM, 1, Edge(4, 2))
x[2]

julia> vanishing_ideal(PM)
Ideal generated by
  -q[2,3,4]^2*q[1,1,1] + q[2,2,1]*q[2,1,2]*q[1,2,2]

julia> coordinate_change(PM)
Ring homomorphism
  from multivariate polynomial ring in 5 variables over QQ
  to multivariate polynomial ring in 5 variables over QQ
defined by
  q[2,3,4] -> 1//3*p[1,2,3] - 1//3*p[1,2,2] - 1//3*p[1,2,1] - 1//3*p[1,1,2] + p[1,1,1]
  q[2,2,1] -> -1//3*p[1,2,3] - 1//3*p[1,2,2] - 1//3*p[1,2,1] + p[1,1,2] + p[1,1,1]
  ...
    \end{minted}

\end{example}

\subsubsection{Phylogenetic Networks} \label{sec:phylo network}

Phylogenetic networks generalize trees to model reticulate evolutionary events such as hybridization, horizontal gene transfer, or recombination. Unlike trees, networks allow internal nodes (called \emph{reticulation} or \emph{hybrid} nodes) to have an in-degree greater than one.
In the algebraic framework, the model parameters include not only the root distribution and transition matrices but also \textit{hybrid parameters} at the reticulation nodes. If a node $v$ has parents $u_1, \dots, u_k$, the state at $v$ is determined by the state at $u_i$ with probability $\lambda_i$, where $\sum \lambda_i = 1$.

\OSCAR supports the construction of phylogenetic networks of level 1 (i.e., networks where each biconnected component contains at most one cycle) and their associated algebraic models using the \texttt{phylogenetic\_network} constructor.

\begin{example}
We construct a Jukes--Cantor model on a \emph{3-leaf sunlet network} (defined as a central 3-cycle with three pendant sink nodes).  As shown in the output, the parameter ring explicitly includes hybrid variables (denoted by \texttt{l}), and the polynomial map \texttt{f} expresses the Fourier coordinates as a convex combination of the underlying tree parametrizations.
\begin{minted}{julia-repl}
julia> E = [[4,1], [5,2], [6,3], [5,4], [6,4], [5,6]];

julia> N = phylogenetic_network(graph_from_edges(Directed, E))
Level-1 phylogenetic network with hybrid nodes {4} and edges
  (4, 1)(5, 2)(5, 4)(5, 6)(6, 3)(6, 4)
  
julia> PMN = jukes_cantor_model(N) # Define a Jukes--Cantor model on the network
Group-based phylogenetic model on a level-1 network with 1 hybrid node, 3 leaves  
  and 6 edges with root distribution [1//4, 1//4, 1//4, 1//4], 
    transition matrices of the form ...
  and fourier parameters of the form [:x, :y, :y, :y].
  
julia> parameter_ring(PMN)[1] # The parameter ring now includes hybrid parameters
Multivariate polynomial ring in 14 variables l[1, 1], l[1, 2], x[1], x[2], ..., y[6]
  over rational field
julia> entry_hybrid_parameter(PMN, Edge(5,4))
l[1, 1]

julia> R, q = model_ring(PMN);

julia> phi = parametrization(PMN);

julia> phi(q[1,1,1])
l[1, 1]*x[1]*x[2]*x[3]*x[4]*x[5] + l[1, 2]*x[1]*x[2]*x[3]*x[5]*x[6]
\end{minted}
\end{example}
This code is part of the broader initiative to create a FAIR database for the \textit{Small Phylogenetic Trees} library \cite{smallTreesPaper}. \Cref{serialization} outlines how we integrate these computational results into a database.

\section{Computing Vanishing Ideals of Parametric Statistical Models}\label{sec:vanishing ideal}
Many core problems in algebraic statistics involve computing the vanishing ideal of the image of a polynomial or rational map (usually in many variables), and thus are fundamentally \emph{implicitization} problems \cite{Sullivant}. The standard technique for solving implicitization tasks is to write down the graph of the rational map (with cleared denominators), then saturate at the denominators and finally to compute a Gröbner basis of the elimination ideal projecting away the parameters. This works for any rational map and only requires the \texttt{parametrization} method to be implemented. However, these default methods often do not terminate on large examples, especially for models with many variables such as phylogenetic networks.  

We implemented specialized techniques which offer immense speed-ups for the problem types which are common in graphical modeling and phylogenetics. Such speed-ups rely on additional structure which are proved by theorems in algebraic statistics. For example, group-based phlogenetic models are toric in their Fourier parameters (see \Cref{sec: group-based}). Knowing this, we rely on the fast tools from the \texttt{4ti2}~package~\cite{4ti2} to compute Gröbner bases. We also incorporate two recently developed methods: \emph{multigraded implicitization}~\cite{MultigradedImp} and \emph{birational implicitization}~\cite{BoegeSolus}. The following examples show how these functions may~be~used and the benefits they reap. All experiments (except where explicitly noted) were run on a machine running Linux~6.12 with an Intel i7-9750H with \SI{32}{\giga\byte} of main memory. Our~comparisons are run with \OSCAR~1.7.0 in \Julia~1.12.4, and \Macaulay{2} version~1.25.11.

\begin{example}
  \label{multi-imp}
  We first show how our software can be used to create a group based Kimura 3 model \cite{Kimura1981} on a four-leaf sunlet network, then we compute up to the degree three components of its vanishing ideal.

\begin{minted}{julia-repl}
julia> G1 = graph_from_edges(Directed,[[5,1],[6,5],[7,6],[7,8],[8,5],[6,2],[7,3],[8,4]]);

julia> N1 = phylogenetic_network(G1)
Level-1 phylogenetic network with hybrid nodes {5} and edges
  (5, 1)(6, 2)(6, 5)(7, 3)(7, 6)(7, 8)(8, 4)(8, 5)

julia> M = kimura3_model(N1)
Group-based phylogenetic model on a level-1 network with 1 hybrid node, 4 leaves  
  and 8 edges with root distribution [1//4, 1//4, 1//4, 1//4], 
    transition matrices of the form 
   [:a :b :c :d;
    :b :a :d :c;
    :c :d :a :b;
    :d :c :b :a]
  and fourier parameters of the form [:x, :y, :z, :t].
julia> phi = parametrization(M)
Ring homomorphism
  from multivariate polynomial ring in 64 variables over QQ
  to multivariate polynomial ring in 34 variables over QQ
defined by
  q[4,4,4,4] -> l[1, 1]*x[6]*t[1]*t[2]*t[3]*t[4]*t[5]*t[7] + ...
  q[4,4,3,3] -> l[1, 1]*x[6]*z[3]*z[4]*z[7]*t[1]*t[2]*t[5] + ...
  ...  

julia> comps = components_of_kernel(3, phi)
Oscar.FinAbGroupElemDict{Vector{QQMPolyRingElem}} with underlying Dict
  Dict{FinGenAbGroupElem, Vector{QQMPolyRingElem}} with 76 entries:
    [3 0 2 1 1 0 2 3 3 0 1 0 1 0 0 0 1 1 0 1 1 1 1 1] => [ ... ]
    [3 0 2 0 1 1 2 3 3 1 1 1 1 1 0 1 0 1 0 0 1 1 0 0] => [ ... ]
    ...

julia> comps[3, 0, 2, 1, 1, 0, 2, 3, 3, 0, 1, 0, 1, 0, 0, 0, 1, 1, 0, 1, 1, 1, 1, 1]
1-element Vector{QQMPolyRingElem}:
 q[4,4,4,4]*q[1,2,1,2]*q[1,1,3,3] - q[4,4,3,3]*q[1,2,1,2]*q[1,1,4,4] - ...
\end{minted}
  
  \Cref{tab:Multigraded} shows timings for our implementation of multigraded implicitization (called \texttt{components\_of\_kernel}) in \OSCAR versus the implementation available in \Macaulay{2} on Kimura~3 networks.

  \begin{table}
  \begin{tabular}{ |c|c|c|c|c| } 
    \hline
    Leaves & Total Degree & Min.\ Gen. & \Macaulay{2} Time (sec) & \OSCAR Time (sec) \\
    \hline
    4 & 2  & 12    & 0.472 & 0.0498  \\ 
    4 & 3  & 64    & 117 & 1.06  \\ 
    5 & 2  & 648   & 329 & 0.951  \\
    5 & 3  & 18560 & - & 372  \\
    \hline
  \end{tabular} \\[.7em]
  \caption{Comparison of the running times between the multigraded implicitization implementations in \Macaulay2 and \OSCAR for the group-based Kimura~3 model on four- and five-leaf sunlet networks.}
  \label{tab:Multigraded}
  \end{table}
\end{example}

\begin{remark}
Part of multigraded implicitization can be parallelized; see \cite{MultigradedImp}.
Since \OSCAR is not thread safe the only chance we have for a parallel implementation is to use multiple processes.
That is, the processes do not share memory and must communicate by serializing (and deserializing) messages. 
This introduces some overhead that depends on the types of the messages and their encodings, often leading to noticeably slower computations in small cases.
For example, the computations in \Cref{tab:Multigraded} 
took orders of magnitude longer when run in parallel.
However, for the three-leaf general Markov model on four states we were able to compute the degree 5 components of the vanishing ideal.
The computation took 2.59 hours running on a machine with an Intel Xeon Silver 4216 using 32 processes and \SI{150}{\giga\byte} of main memory.
This was a large improvement as the non-parallelized time estimate was roughly two days.
For the sake of completeness we show how we setup the computation.
\begin{minted}{julia}
using Oscar
tree = graph_from_edges(Directed,[[4,1],[4,2],[4,3]])
model = general_markov_model(tree)
phi = parametrization(model)
wp = oscar_worker_pool(32)
result = components_of_kernel(5, phi; wp=wp, show_progress=true)
Oscar.close!(wp)
\end{minted}
\end{remark}
The result has been uploaded to zenodo and can be found at \url{https://zenodo.org/records/18312720}; see \Cref{serialization} for more on storing results of computations.

\begin{example}
  Sullivant \cite{GaussianBN} conjectured that the vanishing ideal of a Gaussian DAG model may be computed by saturating its conditional independence ideal at certain principal minors. This conjecture is now a theorem \cite{ColoredDAG} and the proof method was generalized to a ``birational implicitization'' scheme for more classes of models in~\cite{BoegeSolus}. The resulting saturation-based method is the default for Gaussian graphical models in \OSCAR (again leveraging types and multiple dispatch). The elimination-based method is still accesible via the ``algorithm'' keyword argument and below we compare these two methods.
\begin{minted}{julia-repl}
julia> using Oscar, TimerOutputs;
julia> const to = TimerOutput();
julia> E = [[1,3],[1,4],[2,3],[2,4],[4,5], [[i,6] for i in 1:5]...];
julia> M = gaussian_graphical_model(graph_from_edges(Directed, E));
\end{minted}
  \clearpage
\begin{minted}{julia-repl}  
julia> [@timeit to "specialized" vanishing_ideal(M) for _ in 1:20];
julia> [@timeit to "elimination" vanishing_ideal(M; algorithm=:eliminate) for _ in 1:20];
julia> to
────────────────────────────────────────────────────────────────────────
                               Time                    Allocations      
                      ───────────────────────   ────────────────────────
  Tot / % measured:         388s /  73.3%           3.36GiB /   0.9%    

Section       ncalls     time    %tot     avg     alloc    %tot      avg
────────────────────────────────────────────────────────────────────────
elimination       20     284s  100.0%   14.2s   17.5MiB   55.6%   894KiB
specialized       20   62.5ms    0.0%  3.13ms   14.0MiB   44.4%   714KiB
────────────────────────────────────────────────────────────────────────
\end{minted}
The elimination-based method takes on average 14.2 seconds for this example, whereas the specialized routine requires only 3.1 milliseconds while also using slightly less memory. The~same vanishing ideal computation in \Macaulay{2} \cite{M2} done via the \texttt{GraphicalModels} package \cite{GraphicalModelsSource,GraphicalModelsArticle} with \mintinline{macaulay2}{time gaussianVanishingIdeal gaussianRing digraph E} requires 362~seconds.
\end{example}

\section{Serialization}
\label{serialization}
Serialization is the process of converting in-memory data to a format that can be easily stored or transmitted.
Data can often be more valuable than the software that created~it.
For example a research scientist or mathematician may be interested in the result of a computation but may not have the means to run the computation themself.
The ability to serialize results comes with the benefit of not needing to rerun large computations.
We~have implemented serialization for our types allowing the user to store their results in the \texttt{mrdi} file format \cite{FileFormat} available via the \OSCAR \texttt{save} and \texttt{load} functions.
The \texttt{mrdi} format is a JSON-based format \cite{JSON}, which is practical in many ways. First, this makes the file human readable as it is text based as opposed to binary. Secondly, JSON files can be collected into document based databases such as the MongoDB \cite{Mongo} which can provide a helpful querying api.
An example of this is the \texttt{OscarDB}.
Although still in its early stages, it is already accessible via \OSCAR and contains all phylogenetic models from the \url{https://algebraicphylogenetics.org} website~\cite{smalltrees}.
In~future work we plan to have the website pull its data from the \texttt{OscarDB}.
We include a snippet of how one can interface with the \texttt{OscarDB}.

\begin{minted}{julia-repl}
julia> db = Oscar.OscarDB.get_db();

julia> stm = db["AlgebraicStatistics.SmallTreeModels"];

julia> jc_query = Oscar.OscarDB.find(stm, Dict("data.model_type" => "JC"));

julia> jc_models = collect(jc_query)
6-element Vector{Any}:
 Small tree phylogenetic model 3-0-0-JC
 Small tree phylogenetic model 4-0-0-JC
 Small tree phylogenetic model 4-0-1-JC
 Small tree phylogenetic model 5-0-0-JC
 Small tree phylogenetic model 5-0-1-JC
 Small tree phylogenetic model 5-0-2-JC

julia> phylogenetic_model(first(jc_models))
Phylogenetic model on a tree with 3 leaves and 3 edges 
  with root distribution [1//4, 1//4, 1//4, 1//4] and transition matrices of the form 
   [:a :b :b :b;
    :b :a :b :b;
    :b :b :a :b;
    :b :b :b :a]. 

julia> group_based_phylogenetic_model(first(jc_models))
Group-based phylogenetic model on a tree with 3 leaves and 3 edges 
  with root distribution [1//4, 1//4, 1//4, 1//4], 
    transition matrices of the form 
   [:a :b :b :b;
    :b :a :b :b;
    :b :b :a :b;
    :b :b :b :a]
  and fourier parameters of the form [:x, :y, :y, :y].

\end{minted}

Users who have found interesting collections of graphical models can use our serialization methods to populate their own MongoDB or get in touch if they would like to see their collection in the \texttt{OscarDB}.
We also mentioned serialization earlier in \Cref{multi-imp} when discussing communication between separate processes.
With our serialization methods users are free to use our graphical model types across multiple processes if necessary.

\subsection*{Acknowledgements}
  We would like to thank Benjamin Lorenz and Max Horn for their rigorous code review. We would also like to thank Tabea Bacher, Chistiane G\"orgen, Marius Neubert and Matthias Zach for their helpful comments and discussions.
  T.B.\ was funded by the European Union's Horizon 2020 research and innovation programme under the Marie Skłodowska-Curie grant agreement No.~101110545.
  A.D.V\ was funded by the Deutsche Forschungsgemeinschaft (DFG, German Research Foundation) specifically \enquote{MaRDI (Mathematical Research Data Initiative)} (DFG NFDI 29/1, project ID 460135501).
  M.G-L was partially funded by the Beatriu de Pinós postdoctoral programme of the Department of Research and Universities of the Generalitat de Catalunya (ref. 2024BP00235).
  B.H. was partially supported by the Alexander von Humboldt Foundation. 

\bibliographystyle{plain}
\bibliography{biblio}

\end{document}